\documentclass{aastex63}

\received{April 14, 2020}
\revised{May 22, 2020}
\accepted{\today}
\submitjournal{ApJL}

\shorttitle{`Oumuamua was composed of hydrogen ice}
\shortauthors{Seligman and Laughlin}
\graphicspath{{./}{figures/}}

\begin{document}

\title{Evidence that 1I/2017 U1 (`Oumuamua) was composed of molecular hydrogen ice.}

\correspondingauthor{Darryl Seligman}
\email{dzseligman@uchicago.edu}

\author{Darryl Seligman}
\affiliation{ Dept. of the Geophysical Sciences, University of Chicago, Chicago, IL 60637}

\author{ Gregory Laughlin}
\affiliation{Dept. of Astronomy, Yale University,
    New Haven, CT 06517}



\begin{abstract}
`Oumuamua (I1 2017) was the first macroscopic ($l\sim100\,{\rm m}$) body observed to traverse the inner solar system on an unbound hyperbolic orbit. Its light curve displayed strong periodic variation, and it showed no hint of a coma or emission from molecular outgassing. Astrometric measurements indicate that 'Oumuamua experienced non-gravitational acceleration on its outbound trajectory, but energy balance arguments indicate this acceleration is inconsistent with a water ice sublimation-driven jet of the type exhibited by solar system comets. We show that all of `Oumaumua's observed properties can be explained if it contained a significant fraction of molecular hydrogen (H$_{2}$) ice. H$_{2}$ sublimation at a rate proportional to the incident solar flux generates a surface-covering jet that reproduces the observed acceleration. Mass wasting from sublimation leads to monotonic increase in the body axis ratio, explaining `Oumuamua's shape. Back-tracing `Oumuamua's trajectory through the Solar System permits calculation of its mass and aspect ratio prior to encountering the Sun. We show that H$_{2}$-rich bodies plausibly form in the coldest dense cores of Giant Molecular Clouds, where number densities are of order $n\sim10^5$, and temperatures approach the $T=3\,{\rm K}$ background. Post-formation exposure to galactic cosmic rays implies a $\tau \sim 100$ Myr age, explaining the kinematics of `Oumuamua's inbound trajectory.
\end{abstract}



\section{Introduction} \label{sec:intro}

`Oumuamua (1I/2017 U1) passed within 0.25 AU of the Sun prior to its October 19th, 2017 discovery by Robert Weryk \citep{Meech2017} among images from the Pan-STARRS survey. It was initially detected in the course of routine analysis of nightly images taken by the Panoramic Survey Telescope and Rapid Response System (Pan-STARRS) project \citep{Chambers2016}, and following several nights of observation with small telescopes, its extrasolar origin was confirmed, leading to an October 25th, 2017 discovery announcement through the Minor Planet Center \citep{mpec2017a}.  It was then studied intensively over an $l\sim0.13\,{\rm AU}$ segment of its outbound trajectory. The overall geometry of `Oumuamua's encounter with the Solar System is shown in Figure 1.

With eccentricity, $e=1.2$, `Oumuamua encountered the Solar System with $v_{\infty} = 26\,{\rm km\,s^{-1}}$. Its hyperbolic incoming trajectory was consistent with Population I disk kinematics and was remarkably close to the local standard of rest \citep{mamajek2017,bailer-jones2018}. At periastron, `Oumuamua  achieved a heliocentric velocity of 88 ${\rm km\,s^{-1}}$. Despite experiencing irradiation levels $I>20\,{\rm kW\,m^{-2}}$, deep images of `Oumuamua showed no coma \citep{Jewitt2017,Meech2017}, indicating a paucity of $\mu$m-sized dust in its vicinity \citep{Belton2018}. Its spectrum, moreover, displayed no absorption features and skewed red. Its absolute magnitude (based on an average computed from its strongly variable light curve) was measured to be $H\sim22.5$ \citep{Jewitt2017}. If one assumes an albedo $p\sim0.1$, which is typical for small, icy outer Solar System bodies, this implies a size (that is, an effective diameter) of order $d\sim10^4\,{\rm cm}$.

After `Oumuamua's detection and announcement, a variety of observational campaigns were quickly organized on telescopes worldwide, generating a high-quality composite light curve comprising 818 observations and spanning 29.3 days, as summarized by \citet{Belton2018}. Frequency analysis of the light curve shows a power maximum at $P\sim4.3\,{\rm hr}$, which was interpreted to be half the spin period of a rotating body\footnote{Near-uniform rotation is not required to match the light curve. `Oumuamua's rotational motion can exist near the separatrix of a jet-driven pendulum-like Hamiltonian \citep{Seligman2019}}. `Oumuamua's brightness varied by a factor\footnote{The factor of 12 flux variations was computed by averaging the maxima and minima in all oscillations from the digitized composite data.} of $\sim12$, which was explained by positing an elongated shape experiencing complex, non-principle axis rotation \citep{Meech2017,fra18,dra18}. Subsequently, a detailed analysis by \citep{Mashchenko2019} that employed full light curve modeling showed that an oblate $115$:$111$:$19\sim6$:$6$:$1$ ellipsoid provides the most likely geometry for the body. The light curve modeling indicates that a prolate ``cigar-like'' 342 m x 42 m x 42 m (i.e. $a$:$c$:$c\,\sim\,8$:$1$:$1$) ellipsoid also produces a good fit to the data, but requires special tuning of the motion, and is thereby statistically disfavored. 
Observations with the Spitzer telescope described in \citep{Trilling2018} detected no infrared emission from `Oumuamua and placed stringent limits on molecular CO and CO$_2$ (although not H$_2$O) out-gassing.

Extant photometry for `Oumuamua (including multiple HST observations taken through the end of 2017) was compiled and analyzed by \citet{Micheli2018}, who determined that its outbound trajectory was strongly inconsistent with motion subject only to solar gravity. Those authors determined that a radially outward acceleration component of functional form ${ \alpha}=4.92\times10^{-4}(r/1 {\rm AU})^{-2}\,\hat{\bf r}\,{\rm cm\,s^{-2}}$ superimposed on the Keplerian acceleration permits a substantially improved fit to the observed trajectory. The required magnitude of non-gravitational acceleration component, $A_{\rm ng}\sim 2.5\times10^{-4}\,{\rm cm\,s^{-2}}$ at $r\sim 1.4\,{\rm AU}$, where `Oumuamua was observed at highest signal-to-noise, is of order $10^{-3}$ of the solar gravitational acceleration. 

 \citet{Micheli2018} proposed that directed comet-like out-gassing from 'Oumuamua's surface \citep[e.g.][]{Marsden1973} was responsible for the acceleration. A model of this type requires a mass flux of $\dot{m}\sim10^4\,{\rm g\,s^{-1}}$ jetting in the solar direction at $v\sim3\times10^4\,{\rm cm\,s^{-1}}$. This acceleration mechanism is seen in Solar System comets, and the hypothesis that `Oumuamua was a comet-like planetesimal is supported by its featureless red reflection spectrum \citep{Jewitt2017}.
It must be stressed, however, that water's 51 kJ/mol enthalpy of sublimation ensures that driving the acceleration by vaporizing ${\rm H_{2}O}$ ice requires more energy input than `Oumuamua received from solar irradiation \citep{Sekanina2019}. We note that `Oumuamua's non-gravitational acceleration can also potentially be explained by the action of radiation pressure, but this mechanism requires the bulk density to be extremely low, with $\rho < 10^{-5}\,{\rm g\,cm^{-3}}$ \citep[e.g.][]{Loeb2018,moro2019}. In short, 'Oumuamua's acceleration presents a genuine mystery.

\section{Constraints on `Oumuamua's Composition}

Because `Oumuamua's orbit was well-determined, the time-dependent flux of solar energy that it received is known to high accuracy. For models that adopt sublimation-driven out-gassing as the source of the anomalous acceleration, this energy input $E_{tot}=4.5\times10^{13}\,{\rm erg\,cm^{-2}}$ (integrated over the two years surrounding periastron), must provide both sublimation enthalpy as well as the particle kinetic energy of the out-flowing molecules. As was pointed out by \citet{Sekanina2019}, this energy budget imparts a strongly non-trivial constraint and indeed, precludes common molecular species as accelerants. In the case of the current consensus view that posits a prolate geometry for the body and water ice as the substrate, the energy constraint is particularly severe. 

To fix ideas, we can first consider an idealized, one-dimensional model in which a face of a rectangular prism of pure, perfectly absorbing ice is illuminated by normally incident sunlight. The flux, ${\cal N}$, of sublimated molecules leaving a directly illuminated patch of surface ice is
\begin{equation}
{\cal N}=\frac{(1-p)Q(t)-\epsilon \sigma T_S^4}{\Delta H/N_{A}+\gamma kT_S}\, ,
\end{equation}
where $Q(t)$ is the local solar irradiance, $\epsilon$ is the surface emissivity, $\Delta H$ is the sublimation enthalpy of the ice, $T_S$ is the sublimation temperature, $\gamma$ is the adiabatic index of the escaping vapor, and $p$ is the surface albedo  (assumed to be  $p\sim0.1$ throughout), which is controlled by the admixture of impurities in the sublimating ice. 

In the simplest one-dimensional model, a column with unit cross-section, bulk density, $\rho$, and length, $l$ is exposed at one end to the solar flux. For any given volatile species, one finds the length $l$, and the associated mass, $\rho l$, accelerated by sublimation that matches the observed acceleration.

For a species with mass $\mu m_{\rm u}$, the sublimating molecules exit isotropically into the hemisphere associated with the zenith normal to the surface. Evaluating the hemispheric integral, the total number of molecules is twice that necessary to produce the anomalous acceleration with a purely normal exiting outflow. The sublimation velocity is $c_s=\sqrt{ \gamma k T_{S}/(\mu m_{\rm u})}$ \citep{probstein1969}. The outflow thus produces a change in momentum, 
${\cal N}(\mu m_{\rm u})c_s=\delta m2\vert \alpha \vert$, in the unit column, accelerating net mass, $\delta m={\cal N}\,(\mu m_{\rm u} \gamma k T_{S})^{1/2}/2\vert \alpha \vert$ in the anti-solar direction, ${\bf \hat{r}}$. The equivalent length of a column accelerated by a particular species of density $\rho$ is thus given by $l_{e} \sim \delta m/\rho$. For pure H$_2$O ice, using the enthalpy of sublimation listed in Table 1, the equivalent length $l_{e}\sim 15 \,{\rm m}$. This dimension is smaller than the shortest axis length $c=19\,{\rm m}$ of \citet{Mashchenko2019}'s best-fit oblate model, thereby ruling out acceleration arising from water ice sublimation. By contrast, the same calculation for H$_2$ ice (which has a sublimation temperature, $T_{\rm {H}_2}=6\,$K, sublimation enthalpy $\Delta H=1\,{\rm kJ\,mol^{-1}}$, and density, $\rho_{\rm {H}_2}\,{\rm g\, cm^{-3}}$) gives $l_{e}\sim 540\, {\rm m}$, rendering it a far more viable accelerant.

A more accurate assessment requires that the oblate geometry of `Oumuamua be taken explicitly into account. For a given volatile molecule, and a given assumed overall shape, we can calculate the fraction, $f$, of the total surface area that must be covered by exposed ice. The disk-like $a$:$a$:$c\sim6$:$6$:$1$ most-probable model given by \citet{Mashchenko2019} has physical dimensions 115 m x 111 m x 19 m. To simplify analytic calculations, we symmetrize this to 113 m x 113 m x 19 m. In parallel, we can evaluate the 342 m x 42 m x 42 m prolate model.

For these two models, the mass-to-surface-area ratios, $\eta$ are given by

\begin{equation}
 \eta_{obl}=\frac{2c\rho}{3(1+({c^2}/{e a^2})\tanh^{-1}e)}\,,
\end{equation}
and

\begin{equation}
 \eta_{pro}=\frac{2a\rho}{3(1+({a}/{e c})\sin^{-1}e)}\,,
\end{equation}
where $e=\sqrt{1-c^2/a^2}$ is the ellipsoid's eccentricity. 

Aperiodicity in the light curve suggests that `Oumuamua experienced tumbling motion as it traversed the inner Solar System. The total amount of  energy it received from the solar illumination therefore varied in proportion with the instantaneous projected surface area. We define the parameter, $\xi$, which denotes the average projected surface area of the body divided by its total surface area. When isotropically averaged over all viewing angles, $\xi=1/4$ for any convex body \citep{Meltzer1949}, but `Oumuamua's tumbling motion prevented it from receiving fully isotropic time-averaged illumination during its observed acceleration. 

We use the light curve to infer the amount of flux that `Oumuamua received under the assumption of zero solar phase angle and constant surface albedo. With these constraints, the light curve variations correspond to the instantaneous projected surface area reflecting sunlight directly back to the Earth. We assume that the maxima in the light intensity correspond to the maximal projected surface area. For example, for the oblate spheroidal geometry, this would be a head-on view of the disk.  Within these assumptions, the average of the intensity values divided by the maximum intensity gives an estimate of the mean projected surface area as a fraction of the maximum projected surface area.

We introduce the quantity, $\zeta$, which denotes the \textit{maximum} projected surface area as a fraction of the total surface area. For the oblate and prolate geometries, this quantity is given by,

\begin{equation}
 \zeta_{obl}=\frac{1}{2(1+({c^2}/{e a^2})\tanh^{-1}e)}\,,
\end{equation}

and

\begin{equation}
 \zeta_{pro}=\frac{a}{2c(1+({a}/{(e c)})\sin^{-1}e)}\,.
\end{equation}

We then measure the average maxima of each pulse in the digitized light curve, $I_{max}$, and the average intensity $I_{mean}$ over all points, and find that $ I_{mean}/I_{max}\sim0.42$. . This allows us to calculate $\xi$ for both the prolate and oblate geometry, using $\xi = I_{mean}/I_{max}\zeta$. Adopting the 6:6:1 and the 8:1:1 shape models and the high signal-to-noise photometry obtained from 25 Oct 2017 through 1 Nov 2017 \citep{Belton2018}, we find  $\zeta_{oblate}\sim0.47$ and $\zeta_{prolate}\sim0.32$, corresponding to  $\xi_{oblate}\sim0.2$ and $\xi_{prolate}\sim0.13$.

Combining the above relations gives an expression for the  fraction, $f={\eta}/{(\xi\delta m)}$, of an oblate surface that must be covered with sublimating ice to produce the observed radial acceleration,
\begin{equation}
f=\frac{4(\Delta H/N_A+\gamma k T_S)\rho c\vert \alpha(t) \vert}{((1-p)Q(t)-\epsilon \sigma T_S^4)(9\mu m_{\rm u}\gamma k T_S)^{\frac{1}{2}}\xi(1+(\frac{c^2}{ea^2})\tanh^{-1}e)}\,,  \label{eqn:example}
\end{equation}
with an analogous expression holding for the acceleration of a prolate object.

Table 1 lists the covering fractions and other details for a number of species, drawing on the best-fit oblate and prolate models. Sublimating H$_2$ ice can power the jet in both cases. For the oblate geometry, N$_2$, Ne and Ar are also physically viable, but require high covering fractions. Exposed H$_2$ ice needs to cover only a fraction $0.06(\rho/\rho_{\rm H_2})$ of the oblate surface to fully account for the motion.


\begin{table*}
\centering
\caption{Surface covering fraction, $f$, necessary to power a jet consisting of different volatile species. The enthalpy of sublimation ($\Delta$ H),  solid density ($\rho$), and temperature of sublimation ($\textit{T}_{\rm Jet}$) may be found in Table 3 of \citet{Shakeel2018}. We present the surface covering fractions for the best fit oblate  and prolate geometries, denoted by $\textit{f}_{\rm obl}$ and $\textit{f}_{\rm pro}$ respectively. }
\medskip
\begin{tabular}{ccccp{13mm}p{13mm}p{13mm}p{13mm}p{13mm}}
\hline
Species & $\textit{T}_{\rm Jet}$\, [$\rm{K}$] & $\rho$ \, [${\rm g cm^{-3}}$]& $\Delta$ H [${\rm kJ mol^{-1}}$]  & $\textit{f}_{\rm obl}$& $\textit{f}_{\rm pro}$\\
\hline
H$_2$ & 6 & 0.08 & 1 & 0.06& 0.13\\
Ne & 9 & 1.5 & 1.9 & 0.53&1.21\\
N$_2$ & 25 & 1.02 & 7.34 & 0.69&1.58\\
Ar & 30. & 1.75 & 7.79 & 0.97&2.20\\
O$_2$ & 30. & 1.53& 9.26 & 1.09&2.47\\
Kr & 40. & 3.0& 11.53 & 1.45&3.30\\
Xe & 55. & 3.70& 15.79 & 1.69&3.85\\
CO$_2$ & 82. & 1.56& 28.84 & 1.78&4.04\\
H$_2$O & 155. & 0.82& 54.46 & 2.07&4.72\\
\hline
\end{tabular}
\end{table*}

Sublimating ${\rm H}_2$ ice is difficult to observe. With no prominent emission lines, outgassed ${\rm H}_2$ would not have registered in the Spitzer Telescope's 3.6 $\mu {\rm m}$ and 4.5 $\mu {\rm m}$ observations \citep{Silvera1980,Trilling2018}. Moreover, enthalpy-driven cooling associated with the sublimation would have maintained `Oumuamua at close to the $T\sim6\,{\rm K}$ sublimation temperature, obviating detection of purely thermal emission. 

Given that `Oumuamua's dimensions are known with reasonable confidence, and assuming H$_2$ ice sublimation provided the non-gravitational acceleration, the primary remaining unknowns are the bulk density, $\rho$, and the albedo, $p$. `Oumuamua's featureless red reflection spectrum is consistent with the coloration of some Solar System bodies, with close matches in B-V and V-R colors provided by long-period comet nuclei, Trojan asteroids, and active Jupiter-family comets \citep{Jewitt2015}. A tenable conclusion is that `Oumuamua's coloration is generated by a residuum of presolar grain composition -- similar in bulk to the most primitive meteorites -- that remained in increasing concentration as the ${\rm H}_2$ fraction declined. 

\begin{figure*}
\begin{center}
\includegraphics[scale=0.35,angle=0]{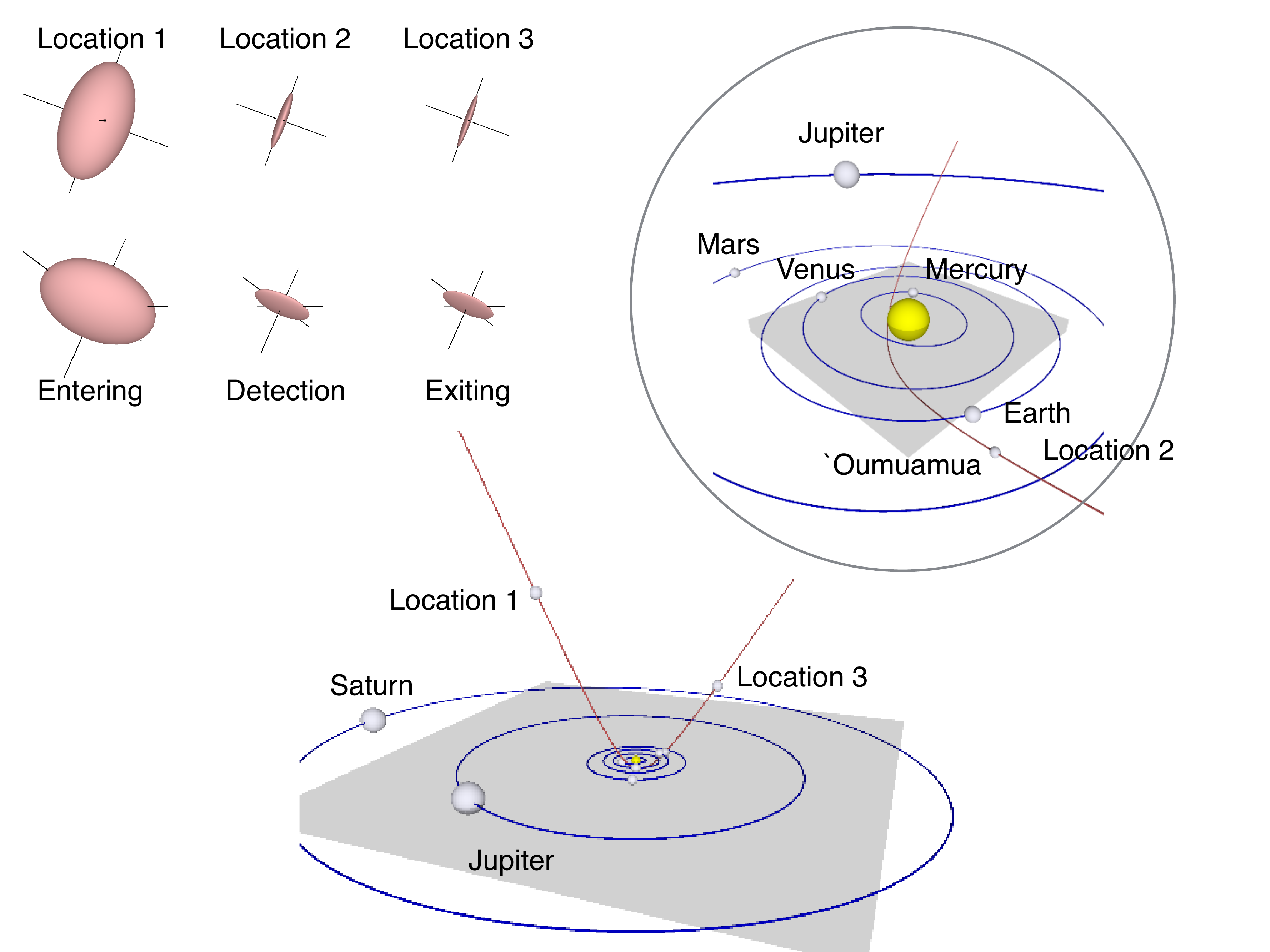} 
\caption{Schematic diagram showing `Oumuamua's size and shape evolution due to H$_2$ sublimation and its trajectory through the Solar System. Pairs of orientations at three discrete points on the trajectory are shown in the upper left. }
\end{center}
\end{figure*}
\begin{figure*}
\begin{center}
\includegraphics[scale=0.55,angle=0]{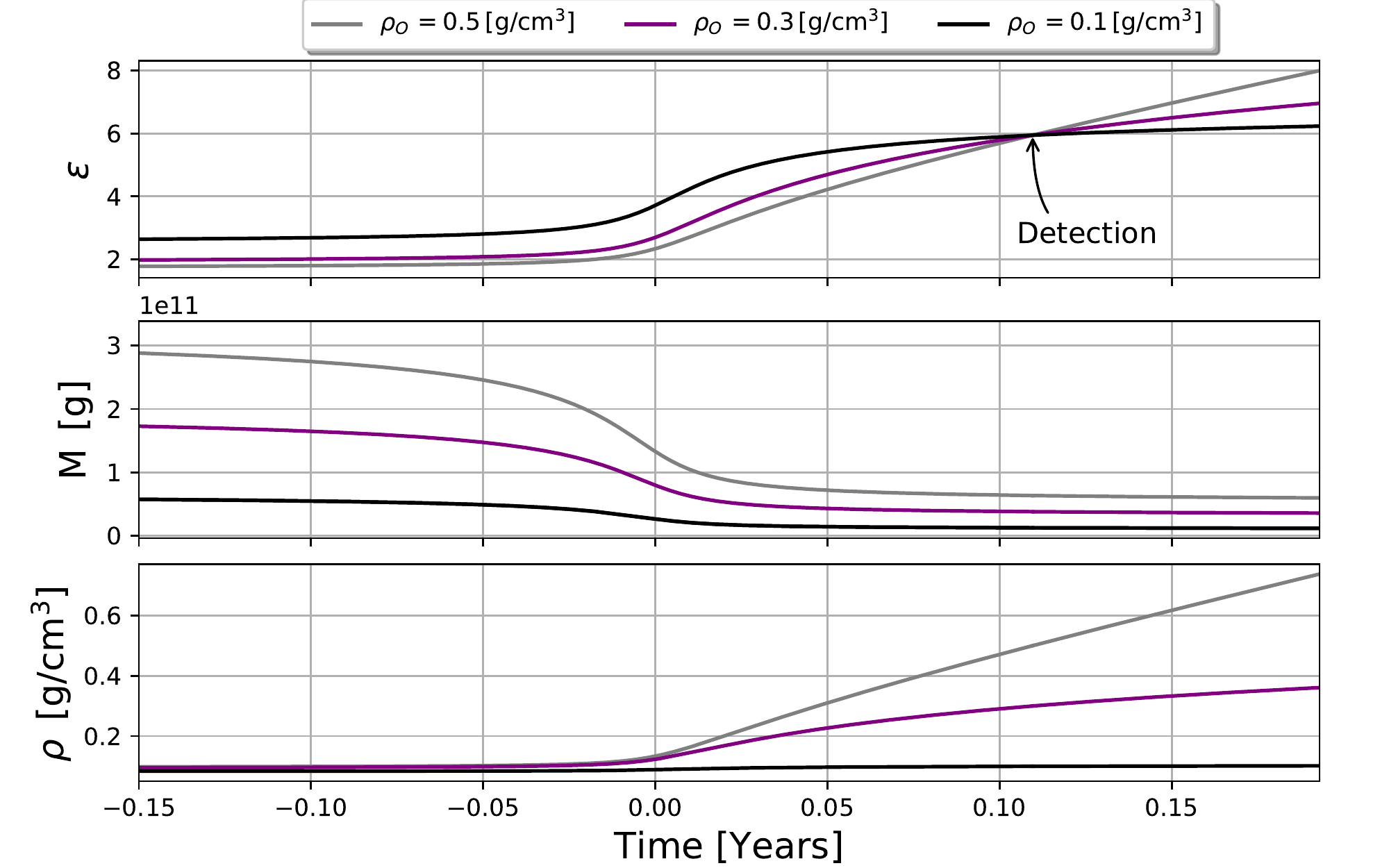} 
\caption{ Density, mass and aspect ratio evolution for `Oumuamua as it travelled through the Solar System. The periastron passage is displayed at a time, $t=0$,  and the date of detection (Location 2 in Figure 1) is labeled. }
\end{center}
\end{figure*}

\section{Evolution of 'Oumuamua's shape}

Soon after 'Oumuamua's discovery,  \citet{Domokos2017} suggested that its unusual shape resulted from abrasion by small particles. The process of isotropized photon-driven sublimation will also produce an evolution that conforms to \citet{Domokos2017}'s theoretical model. Sublimation driven by uniform illumination generates a secular increase in the aspect ratio of the body, with large ratios being reached after a majority of the original mass has been removed. 

We have incorporated this phenomenon in a semi-analytic model that back-traces `Oumuamua's motion through the Solar System and recovers the properties with which it entered the system. We assume that `Oumuamua continuously sublimated H$_2$ at the rate required to generate the observed $\alpha \propto 1/r^{2}$ non-gravitational acceleration, and we assume that `Oumuamua's tumbling motion isotropized exposure of its surface to the Sun over a time scale ($\sim$days) that is shorter than the time scale over which its received flux varied significantly ($\sim$weeks). Although the degree of tumbling may not have been sufficient to fully isotropize the exposure, this assumption permits assessment of the density, size and aspect ratio evolution that `Oumuamua experienced while in the vicinity of the Sun.

With our timescale ordering, H$_2$ gas exits normally from the surface in thin ellipsoidal shells. In each small time increment, the geometry of `Oumuamua shifts (approximately) into a new ellipsoid whose axes are each increased or decreased by an amount $\delta h$. We simulate this process using time-stepping. We start at the moment of the first observation of `Oumuamua in October 2017, and integrate both forward and backward in time.

Mass loss at each time step is governed by the strength of the non-gravitational acceleration. We assume that the best-fit functional form, $\alpha(r)$ is sustained throughout the trajectory. We calculate the change in mass, $\delta m$, at each time-step of length $\Delta t$, and position in the orbit, $r$, by invoking the conservation of momentum, where

\begin{equation}
    \delta m = \frac{2 \alpha(r)\rho_{\rm{bulk}}(t)V(t)\Delta t}{\sqrt{ \gamma k T_{S}/(\mu m_{\rm u})}}\label{deltameq}\, .
\end{equation}
Here, $V(t)$ and $\rho_{\rm{bulk}}(t)$ are the volume and  bulk density at each time-step. The resulting differential volume shell,  $\delta V$, is given by $ \delta V =\delta m/\rho_{H_2}$. In order to  explicitly demand the conservation of mass and ellipsoidal geometry, we find the root of the cubic equation, $p(\delta h)$ that satisfies,

\begin{equation}
     p(\delta h)=\delta V-\frac43\pi (a\pm\delta h)(b\pm\delta h)(c\pm\delta h)=0\,,
\end{equation}
where we use the $\pm$ solutions for integrating backwards and forwards in time respectively. Since this differential volume shell is assumed to consist of pure molecular hydrogen, we calculate the resulting bulk density of `Oumuamua at each timestep using,

\begin{equation}
     \rho_{\rm{bulk}}(t\pm\Delta t)=\frac{\rho_{\rm{bulk}}(t)V(t)\mp \delta m}{V(t\pm\Delta t)}\ .
\end{equation}
 We verified that the numerical scheme had converged in timesteps, strictly demanded mass conservation and produced results that were time reversible. This description is, of course, idealized. In reality the outflow may have other molecules as well as dust entrained within it (although the observations do provide strict upper limits). Accounting for constituents other than H$_2$ would result in a coeffiecient of order unity in Equation \ref{deltameq}.

The evolution of the body in response to the time-varying solar irradiation is displayed in Figure 2. We show simulations that start with bulk densities at the date of detection of $\rho_O=0.1$, $0.3$ and $0.5 \, \rm{g \,cm^{-3}}$. We chart the evolution over a time period of $\tau\sim 3$ months encompassing the most dramatic evolution, but we evolve the simulation for $\tau\sim 2$ years centered on $t=0$ corresponding to the periastron passage. The behavior of the aspect ratio, size and density evolves asymptotically in the regions of the simulation not displayed in the plot. As an example, if 'Oumuamua had a density of $\rho_O= 0.3 \, \rm{g \,cm^{-3}}$ at the time when it was first observed, its dimensions shrank from $\sim$196 m x 196 m x 102 (with $\rho_{\rm{bulk}}\sim0.093 \, \rm{g \,cm^{-3}}$)  upon entering the solar system to
$\sim$113 m x 113 m x 19 m when observed to $\sim$110 m x 110 m x 16 m upon exiting (with $\rho_{\rm{bulk}}\sim0.36 \, \rm{g \,cm^{-3}}$). As a consequence of the encounter with the Sun, the total mass dropped by a factor of 5 -- from $2\times10^{11}\,{\rm g}$ to $4\times10^{10}\,{\rm g}$. (A visual depiction of this evolution is presented in Figure 1). We conclude that, for a range of plausible initial conditions, `Oumuamua entered the Solar System with an aspect ratio $\epsilon\sim2-3$ and a bulk density $\rho_{\rm{bulk}}\sim0.1\,\rm{g \,cm^{-3}}$.   \footnote{The script to generate the data in Table 1, these simulations and Figure 2 may be found at \url{https://github.com/DSeligman/Oumuamua_Hydrogen}.}

We can extend the calculation further backward in time to investigate the geometric evolution as `Oumuamua traveled through the galaxy and before it encountered the Solar System. We perform a similar simulation tracking the aspect ratio, size and density as  it was exposed to the galactic cosmic ray flux $\Phi_{CR}\sim 10^9\, {\rm ev\, cm^{-2} s^{-1}}$ \citep{White1996}. Using the initial conditions from the preceding calculation for the case where $\rho_O\sim0.3 \, \rm{g \,cm^{-3}}$, we calculate that `Oumuamua reached primordial aspect ratios of $\epsilon \sim 1.74$, $\sim 1.54$ and $ \sim 1.28$ after  $\tau \sim 10\, \rm{Myr}$,  $ \sim 30\, \rm{Myr}$ and  $ \sim 100\, \rm{Myr}$ of evolution,  respectively, with dimensions of $\sim$220 m x 220 m x 130 m, $\sim$270 m x 270 m x 180 m, and $\sim$440 m x 440 m x 340m.  We also extended the calculation forward in time, and found that if it had not encountered the Solar System, `Oumuamua would have survived for an additional $ \sim 37\, \rm{Myr}$ before being reduced to a remnant with bulk density of $\rho_{\rm{bulk}}\sim0.8\,\rm{g \,cm^{-3}}$. In this simple model, prior to encountering the Solar System, ‘Oumuamua still had an appreciable fraction of it’s estimated total lifetime remaining.

\section{Formation of objects rich in H$_2$ ice}
${\rm H}_2$ in the gas phase forms via dust-catalyzed reactions and is the dominant constituent of Giant Molecular Clouds [GMCs] \citep{Hollenbach1971, Wakelam2017}. The coldest, highest-density regions within GMCs are prestellar cores, which concentrate along filaments that pervade the overall cloud structure and have $T\le10\,{\rm K}$  and number densities, $n\ge10^5 - 10^{6}\,{\rm cm}^{-3}$ \citep{Andre2014}. At the 2.7K cosmic background temperature, solid ${\rm H}_2$ has a sublimation vapor pressure corresponding to $n\simeq 3\times10^{5}\,{\rm cm}^{-3}$\citep{Anderson1989}. In regions of low cosmic ray density (such as in the inner regions of M31), CO observations suggest that temperatures very close to the microwave background can be achieved \citep{Laurent1998}. Recent observations by \citet{Kong2016} of line emission from multiple transitions involving N$_2$D$^+$ and N$_2$H$^+$ in massive starless/early-stage molecular cloud cores points to excitation temperatures of $T\sim4\,{\rm K}$. Identification of even colder gas may be possible through observation of ortho-H$_2$D$^+$, which has been detected with the Atacama Large Millimeter/submillimeter Array \citep{Friesen2014}. Given a temperature of order 3$K$ in a dense core, nucleated growth of solid hydrogen onto interstellar grains can occur. 

Typical interstellar grains, formed from carbon or silicon compounds have sizes, $D_{\rm g}\sim1{\mu}{\rm m}$, cross sectional areas, $\sigma_{\rm g}\sim10^{-8}\,{\rm cm^2}$, and masses $m_{\rm g}\sim10^{-12}\,{\rm g}$ \citep{Draine2003}. For a dense molecular cloud core with $n_{\rm H2}=10^{6}\,{\rm cm^{-3}}$ and metallicity $Z=0.01$, this implies a grain number density $n_{\rm g}=3\times10^{-8}\,{\rm cm}^{-3}$. Equipartition at $T=2.7\,{\rm K}$ gives a grain velocity $v_{\rm g}\sim0.03\,{\rm cm\,s^{-1}}$ (implying a grain-grain collision rate of $\Gamma=n_{\rm g}\sigma_{\rm g} v_{\rm g}=10^{-17}\,{\rm s}^{-1}$, which is entirely negligible). For $n_{\rm H2}=10^{6}\,{\rm cm^{-3}}$ and $T=2.7\,$K, the sublimation vapor pressure is exceeded by a factor of 3, and molecular hydrogen will freeze out onto the interstellar grains. This process is relatively rapid. The collision rate of ${\rm H}_2$ molecules onto grains leads to a grain mass increase rate $dM_{\rm g}/dt\sim6\times10^{-22}\,{\rm g\,s^{-1}}$,  and growth timescale $\tau\sim 500\,{\rm yr}$ (for sticking probability, $s=0.1$). Hence, within several times $10^{4}$ years of the onset of freeze-out, a significant fraction of the ${\rm H}_2$ will sequester onto the small grains.

When condensation conditions for ${\rm H}_2$ are achieved,  the resulting H$_2$-mantled grains in a GMC core are subject to several charging processes which aid the growth of large dust aggregates in regions where the gas number density is in the range $10^{4}\,{\rm cm}^{-3} < n < 10^{6}\,{\rm cm}^{-3}$ \citep{Ivlev2015}. In short, cold plasma charging tends to give grains negative charge; cosmic rays ionize gas molecules, and electrons, due to their higher velocities, preferentially land on grains. On the other hand, UV radiation stemming from the cosmic ray ionization events produces photoelectric charging of grains which imparts positive charge to grains. \citet{Ivlev2015} show that competition between the cold-plasma collection and photo-emission will create  approximately equal abundances of positively and negatively charged dust. Critically, their derived optimum is size-independent, permitting large aggregates to be produced. An H$_2$-rich `Oumuamua-precursor with $r=300\,{\rm m}$ requires H$_2$ ice condensates to coagulate  within a $D\sim10^5\,{\rm km}$ sided volume of the GMC. If this is to occur over a $\tau\sim10^4$ yr time scale, characteristic grain velocities  of order $v\sim3\times10^{-2}$ cm/s are required.

Overall, star formation in GMCs is of order $1\%$ efficient \citep{Shu1987}, with higher efficiencies observed for the densest core regions \citep{Konyves2015}. Macroscopic bodies composed of frozen molecular hydrogen that are not incorporated into stellar systems will be released into low-velocity dispersion galactic orbits, with survival times determined by the ambient cosmic ray flux.  With the backtracing calculation described above, we estimate `Oumuamua's initial mass, radius, and age are roughly $3\times10^{12}\, {\rm g}$, $220$ m and $10^8$ years respectively if its primordial aspect ratio was of order $ \sim 1.3$, and if its density at the start of observations was $\rho_0=0.3\,{\rm g\,cm^{-3}}$.

\section{Future Observations}

If `Oumuamua's anomalous acceleration stemmed from sublimating H$_2$ ice, it is likely that a large population of similar objects exist. An analysis by \citet{Do2018} suggests that the space density of `Oumuamua-like objects is $n=0.2\,{\rm AU^{-3}}$. Our estimate of `Oumuamua’s initial mass thus suggests a total mass of $\sim1\,M_{\oplus}$ of H$_2$-rich bodies per star. A galactic sea of unbound planetesimal-sized objects has potential consequences for star and planet formation \citep{Bannister2019}, and population members will be readily detectable with the forthcoming Large Synoptic Survey Telescope. ESO’s proposed \textit{Comet Interceptor} \citep{jones2019} mission, moreover, will be well-positioned to provide \textit{in situ} studies \citep{Seligman2018}.

\acknowledgements
 We thank Fred Adams, Steve Desch, Shuo Kong and Scott Sandford for useful conversations. 


\bibliography{sample63}{}
\bibliographystyle{aasjournal}


\end{document}